\documentclass[lettersize,journal]{IEEEtran}
\usepackage{amsmath,amsfonts}
\usepackage{algorithmic}
\usepackage{algorithm}
\usepackage{array}
\usepackage[caption=false]{subfig}
\usepackage{textcomp}
\usepackage{stfloats}
\usepackage{url}
\usepackage{verbatim}
\usepackage{graphicx}
\usepackage{cite}
\graphicspath{{Fig/}}
\usepackage{color} 
\usepackage{multirow}
\usepackage{hyperref}

\begin{document}

\title{No-­reference Point Cloud Geometry Quality Assessment Based on Pairwise Rank Learning}

\author{Zhiyong~Su,
	Chao~Chu,
	Long~Chen,
	Yong~Li,
	and~Weiqing~Li
	
	\thanks{Manuscript received 00 00, 0000; revised 00 00, 0000.  This work was supported in part by the National Key R\&D Program of China under Grant 2018YFB1004904. (Corresponding author: Zhiyong Su.)}
	
	\thanks{Zhiyong Su, Chao Chu, Long Chen, and Yong Li are with the School of Automation, Nanjing University of Science and Technology, Nanjing, Jiangsu Province 210094, P.R. China (e-mail: su@njust.edu.cn, 979953220@qq.com, 2712371060@qq.com, 2422652020@qq.com)} 
	
	\thanks{Weiqing Li is with the School of Computer Science and Engineering, Nanjing University of Science and Technology, Nanjing, Jiangsu Province 210094, P.R. China (e-mail: li\_weiqing@njust.edu.cn).}}

\markboth{Journal of \LaTeX\ Class Files,~Vol.~14, No.~8, August~2021}%
{Shell \MakeLowercase{\textit{et al.}}: A Sample Article Using IEEEtran.cls for IEEE Journals}


\maketitle

\begin{abstract}
Objective geometry quality assessment of point clouds is essential to evaluate the performance of a wide range of point cloud-based solutions, such as denoising, simplification, reconstruction, and watermarking.
Existing point cloud quality assessment (PCQA) methods dedicate to assigning absolute quality scores to distorted point clouds.
Their performance is strongly reliant on the quality and quantity of subjective ground-truth scores for training, which are challenging to gather and have been shown to be imprecise, biased, and inconsistent.
Furthermore, the majority of existing objective geometry quality assessment approaches are carried out by full-reference traditional metrics.
So far, point-based no-reference geometry-only quality assessment techniques have not yet been investigated.
This paper presents PRL-GQA, the first pairwise learning framework for no-reference geometry-only quality assessment of point clouds, to the best of our knowledge.
The proposed PRL-GQA framework employs a siamese deep architecture, which takes as input a pair of point clouds and outputs their rank order.
Each siamese architecture branch is a geometry quality assessment network (GQANet), which is designed to extract multi-scale quality-aware geometric features and output a quality index for the input point cloud.
Then, based on the predicted quality indexes, a pairwise rank learning module is introduced to rank the relative quality of a pair of degraded point clouds.
To train the proposed PRL-GQA framework, a new rank dataset named PRLD is constructed, which includes 150 reference point clouds and 15750 pairs of distorted samples.
In addition, a large-scale quality-annotated dataset containing 5250 geometrically distorted samples with pseudo-MOS is also established for fine-tuning the pre-trained GQANet to predict absolute quality scores.
Extensive experiments demonstrate the effectiveness of the proposed PRL-GQA framework.
Furthermore, the results also show that the fine-tuned no-reference GQANet performs competitively when compared to existing full-reference geometry quality assessment metrics. 
The source code and datasets will be publicly available at: \href{https://zhiyongsu.github.io/Project/PRLGQA.html}{https://zhiyongsu.github.io/Project/PRLGQA.html}.
\end{abstract}


\section{Introduction}
\label{sec:introduction}
\IEEEPARstart{N}{owadays}, the rapid development of geometric sensing techniques (e.g. time-of-flight range finding and structural lighting)  have witnessed the widespread use of point clouds in various fields, such as 3D reconstruction, autonomous driving, and augmented reality.
Unfortunately, point clouds always suffer from more or less degradation during their acquisition, coding, simplification, denoising, watermarking, etc \cite{wuxj21}.
Therefore, reliable geometry quality assessment of point clouds is essential and critical, since such quality metrics can be used in the design and optimization of point cloud-based algorithms and systems.
Take the point cloud denoising for example, objective geometry quality metrics are usually employed to evaluate the geometry quality of denoised point clouds, followed by optimizing the denoising process or tuning denoising parameters.
These kinds of algorithms and systems mainly concentrate on how to evaluate the geometry-only quality objectively, without the consideration of color or texture information. 

Naturally, the geometry quality of point clouds can be evaluated by either subjective assessment methods or objective assessment methods.
Subjective quality assessment is a straightforward and reliable way to evaluate the geometry quality of point clouds. 
And, it is also the most convenient way to generate a ground truth for judging the performance of objective assessment methods \cite{Nehmey19}.
However, despite its importance, subjective quality assessment is time-consuming and cumbersome, thereby cannot be embedded into real-time point cloud processing applications.
In contrast, objective quality assessment, which is still an open problem, can accurately predict the level of impairment of degraded point clouds effectively.
Up to now, a number of objective point cloud quality assessment (PCQA) methods that work well for different types of distortions and visualization approaches have been reported \cite{Alexioue18,Fretesh19,Nehmey19,niuyz19,Yangq21,liuq22,Freitasxg22}.
Despite the considerable progress made by existing methods, there are still several limitations in objective geometry quality assessment.

First of all, existing PCQA methods devote to assigning absolute quality scores to degraded point clouds.
Thus, their performance depends heavily on the quality and quantity of ground-truth scores for training, such as the mean opinion score (MOS),  which is the average of quality scores given by multiple observers.
However, it is observed that it is much easier for people to rate the relative quality of two given point clouds than to assign absolute quality scores to them.
Therefore, subjective quality scores are usually imprecise, biased, and inconsistent \cite{Gaof15,mkd17,Nehmey19,Hub19,oufz21}.

Second, the majority of existing objective geometry quality assessment metrics are mainly carried out by full-reference traditional approaches, in which original reference point clouds are always needed.
Current deep learning-based methods all target to both geometry and color distortions \cite{wuxj21,Yangq21}, which can not be directly used for geometry-only quality assessment.
Furthermore, most of these learning-based methods adopt a projection-based assessment strategy that projects 3D point clouds into a set of 2D planes.
However, projection-based methods inevitably suffer from information loss and cannot characterize the 3D points distribution very well.

Third, existing PCQA datasets contain a limited number of samples, such as IRPC \cite{Javaheri21}, CPCD 2.0 \cite{Hual21}, SJTU-PCQA \cite{Yangq21}, SIAT-PCQD \cite{wuxj21}, and WPC \cite{liuq22}.
In these datasets, most of the reference point clouds are colored point clouds.
And, their ground-truth scores are obtained by considering both the geometry and color information.
Therefore, it is difficult to directly employ existing datasets to derive and train learning-based metrics with high generalization ability for the geometry-only quality assessment.
And, it is also extremely labor-intensive and expensive to collect and annotate sufficient samples to build qualified datasets.

To address the above limitations, this paper presents a no-reference point-based geometry-only quality assessment framework based on deep pairwise rank learning, named PRL-GQA, without the need of reference point clouds which are not always available in real scenarios.
The proposed PRL-GQA framework takes a pair of degraded point clouds as input, and leverages the pairwise rank learning to predict their relative geometry quality order.
To this end, the PRL-GQA, which takes advantage of both rank learning and quality score prediction, consists of a siamese architecture and a pairwise rank learning module.
Each branch of siamese architecture is a geometry quality assessment network (GQANet), which is designed to calculate a quality index for the input point cloud.
The pairwise rank learning module learns to rank the relative quality of a pair of degraded point clouds based on predicted quality indexes.
To train the proposed PRL-GQA framework, a pairwise ranking dataset is constructed, which provides the quality ranking order of a pair of point clouds as ground truth.
Furthermore, the pre-trained GQANet can also be employed to predict absolute quality scores after fine-tuning on a newly built large-scale quality-annotated dataset with pseudo-MOS.

The main contributions of this paper are as follows:
\begin{enumerate}
	\item[1)] A pairwise rank learning framework for no-reference point cloud geometry-only quality assessment (PRL-GQA) is proposed, in which the geometric feature extraction, weighted quality index calculation, and pairwise rank learning are jointly optimized in an end-to-end manner.
	To the best of our knowledge, this is the first deep pairwise ranking learning-based geometry-only quality assessment method. 
	
	\item[2)] A geometry quality assessment network (GQANet) is designed to extract multi-scale patch-wise quality-aware geometric features and predict both the quality index as well as the weight of each patch directly in the 3D space rather than 2D image space.
	The pre-trained GQANet can also be employed to predict absolute quality scores after fine-tuning on datasets with ground-truth scores.

	\item[3)] A large-scale ranked dataset called PRLD, containing thousands of geometrically distorted point cloud pairs, is constructed to train the proposed PRL-GQA. 
	
	\item[4)] A large-scale quality-annotated dataset with pseudo-MOS, named PCGD-PMOS, is established to fine-tune the pre-trained GQANet to predict absolute quality scores.
	
\end{enumerate}

The rest of this paper is organized as follows. 
Section \ref{sec:Related Works} reviews the studies about PCQA. 
Section \ref{sec:Proposed Method} describes the proposed geometry quality assessment framework based on pairwise learning in detail. 
Section \ref{sec:Dataset} introduces the construction of PRLD and PCGD-PMOS datasets. 
Section \ref{sec:Experimental} presents experimental studies to demonstrate the state-of-the-art performance of the proposed framework. 
Section \ref{sec:Conclusion} concludes the paper.

\section{Related Works}
\label{sec:Related Works}


This section reviews objective PCQA approaches, which may be categorized into point-based and projection-based methods from the perspective of feature extraction.
Besides, according to their reliance on the availability of an original reference point cloud, objective quality measures can also be divided into three categories: full-reference, reduced-reference, and no-reference \cite{Alexioue18,Dumice18}.
Full-reference (FR) methods compare a degraded point cloud to the original reference point cloud.
Reduced-reference (RR) approaches require information derived from the original point cloud.
And, no-reference (NR) techniques operate solely on the degraded point cloud. 

\subsection{Point-based methods}
\label{subsec:Point-based methods}

Point-based quality assessment methods evaluate the quality of point clouds based on quality-related attributes in the 3D space directly, such as geometry and color information.

Up to now, point-based methods mainly carry out by full-reference metrics \cite{Alexioue18,Dumice18,Meynetg20,Yangq22,Dinizr21,liuq22}.
The majority of these methods focus on geometric distortions, which can be distinguished into point-to-point, point-to-plane, point-to-distribution, and plane-to-plane metrics.
The point-to-point metric computes the distance between points in the degraded point cloud and their corresponding points in the reference point cloud, such as Hausdorff Distance (HD) \cite{Lavoueg10}, Root Mean Square Error (RMSE) \cite{Javaheria17a}, Mean City-block Distance (MCD) \cite{Zengj20}, Peak signal to noise ratio (PSNR) \cite{Tiand17}, and Chamfer distance (CD) \cite{Zhangdb21}.
The point-to-plane metric is calculated by projecting the vector that connects two associated points along the normal vector of the reference point \cite{Tiand17}.
The point-to-distribution distance adopts a new type of correspondence between two point clouds and employs the Mahalanobis distance to measure the distance between a point and a distribution of points from a small point cloud region \cite{Javaheria20a}.
The plane-to-plane metric is built on the angular similarity of tangent planes that correspond to associated points between the reference and the degraded point cloud \cite{Alexioue18icme}.

Besides geometry information, color information also plays an important role in the point cloud visual quality assessment.
Viola et al. extracted color statistics from both reference and degraded point clouds, and employed color histograms to drive objective metrics \cite{Violai20}.
Moreover, they combined color-based and geometry-based metrics to provide a global quality score.
Meynet et al. also selected several geometry-based and color-based features and combined them linearly by logistic regression \cite{Meynetg20}.
In \cite{Yangq22}, Yang et al. proposed a metric, called GraphSIM, to predict the human perception of colored point clouds with superimposed geometry and color impairments.
They employed graph signal processing to extract point cloud color gradient to yield robust quality prediction. 
Diniz et al. developed a BitDance metric that uses color and geometry texture descriptors \cite{Dinizr21}.
They compared and combined the statistics of color and geometry information of the reference and original point cloud to estimate the perceived quality of the degraded point cloud.
The proposed CPC-GSCT metric employed geometric segmentation and color transformation respectively to construct geometric and color features to estimate the point cloud quality \cite{Hual21}.

No-reference PCQA methods only utilize degraded point clouds and do not require information about reference point clouds.
However, currently, there is less research on no-reference quality assessment in the 3D space directly.
The proposed BQE-CVP method employed handcraft features to develop a no-reference quality evaluator \cite{Hual21cvp}.  
It characterized the distortion of distorted point clouds from geometric, color, and joint perspectives.
Liu et al. proposed a no-reference metric ResSCNN based on a sparse convolutional neural network to accurately estimate the subjective quality of point clouds \cite{Liuyp22}.
The proposed ResSCNN adopts a hierarchical feature extraction module to extract both geometry and color attributes of point clouds, which takes the entire point cloud as input.

In a word, existing point-based approaches are all designed to assign absolute quality scores to degraded point clouds.
And, these methods mainly concentrate on full-reference metrics, in which reference point clouds are usually unavailable in some practical applications.
Most important of all, existing no-reference PCQA metrics are all projection-based approaches \cite{Yangq22,Liuyp22}.
Deep learning-based geometry-only quality assessment techniques in the 3D space have not been explored so far.

\subsection{Projection-based methods}
\label{subsec:Projection-based methods}

Projection-based methods dedicate to projecting 3D point clouds into a set of 2D planes and then inherit existing image quality assessment methods to assess the quality of point clouds.
Currently, projection-based approaches have dominated the field of PCQA, which take both geometry and color information into consideration.

Most existing projection-based methods employ handcraft features to characterize the distortion of degraded point clouds.
Alexiou et al. investigated the impact of the number of viewpoints employed to assess the visual quality of point clouds and proposed to assign weights to the projected views based on interactivity information obtained during subjective evaluation experiments \cite{Alexioue19mex}.
In \cite{wuxj21}, using a Head-Mounted Display (HMD) with six degrees of freedom, Wu et al. proposed two projection-based objective quality evaluation methods: a weighted view projection based model and a patch projection-based model. 
He et al. projected the texture information and curvature of colored point clouds onto 2D planes, and extracted texture and geometric statistical features, respectively.
Their method combined both colored texture and curvature projection.
Yang et al. projected the 3D point cloud onto six perpendicular image planes of a cube to obtain both 2D texture and depth images to represent the photometric and geometric information of the original point cloud \cite{Yangq21}.
Then, they utilized features extracted from these images for objective metric development.
Liu et al. employed an attention mechanism and a variant of information content-weighted structural similarity to develop a novel objective model, which significantly outperforms existing metrics. 

Recently, deep learning-based feature extraction methods have also been introduced into projection-based approaches.
Liu et al. proposed a deep learning-based no reference PCQA method, called PQA-Net, which consists of three modules: a multi-view-based joint feature extraction and fusion (MVFEF) module, a distortion type identification (DTI) module, and a quality vector prediction (QVP) module \cite{liuq21tcsvt}.
The entire network is jointly trained for quality prediction. 
Tao et al. projected 3D point clouds into 2D color projection maps and geometric projection maps and designed a multi-scale feature fusion network  to blindly evaluate the visual quality \cite{Taowx21}. 

All in all, existing projection-based methods, like point-based approaches, dedicate to predicting absolute quality scores for input distorted point clouds as well.
However, the 3D-to-2D projection cannot characterize the 3D points distribution very well and inevitably result in information loss.
Meanwhile,  the selection of projection directions may significantly influence the overall assessment performance \cite{Yangq22,Liuyp22}.
Therefore, performing point cloud geometry quality assessment directly in the 3D space is a promising and challenging task.
This paper dedicates to developing a point-based deep pairwise rank learning framework for no-reference point cloud geometry-only quality assessment.

\section{Proposed Method}
\label{sec:Proposed Method}

\subsection{Overview}
\label{subsec:Overview}

\begin{figure*}[!htbp]
	\centering
	\includegraphics[width=1.0\textwidth]{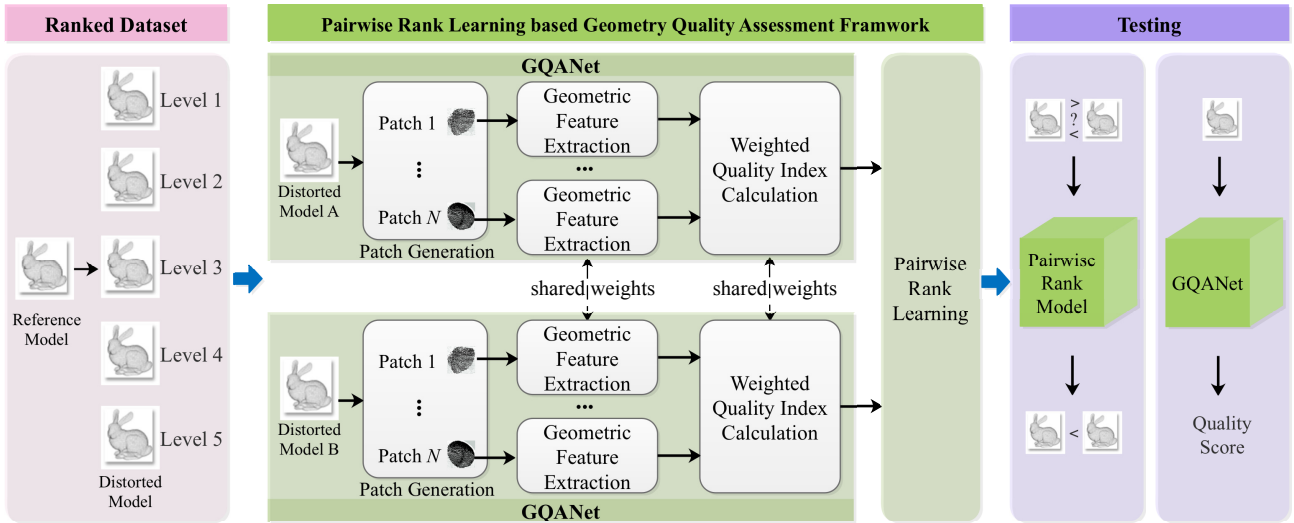}
	\caption{Overview of the proposed PRL-GQA framework. In the training phase, given two pairwise ranked point clouds, the GQANet first extracts patch-wise quality-aware geometric features and then calculates the quality index and weight of each patch. The overall quality indexes of a pair of input point clouds are passed to the pairwise rank learning module. In the testing phase, the proposed PRL-GQA framework can be employed to rank two degraded point clouds. Meanwhile, the pre-trained GQANet can also be used to predict quality scores after fine-tuning.}
	\label{fig:Overview}
\end{figure*}

The proposed pairwise rank learning-based no-reference point cloud geometry-only quality evaluation framework (PRL-GQA) is composed of a siamese network and a pairwise rank learning module, as illustrated in Fig. \ref{fig:Overview}.

The siamese network contains two identical branches, named geometry quality assessment network (GQANet) which comprises patch generation, geometric feature extraction, and weighted quality index calculation.
The patch generation module first segments the input degraded point cloud into different overlapping patches.
Then, a geometric feature extraction module is introduced to extract patch-wise quality-aware geometric features of each distorted point cloud, which can well capture point features in multiple scales. 
Finally, a weighted quality index calculation module is specifically designed to predict both the quality index and the weight of each patch.
These indexes and weights of all patches are finally combined to calculate the quality index of the input point cloud.

The pairwise rank learning module determines the relative quality probability label according to the quality index predicted by the two branches (GQANet).
The learned PRL-GQA framework can rank point clouds according to their distortion levels.
Furthermore, it is also able to employ the pre-trained GQANet to score distorted point clouds after fine-tuning on datasets with ground-truth scores.

\subsection{Patch Generation}

The patch generation partitions the input point cloud into a fixed number of overlapping patches (i.e. $N$) with a fixed number of neighboring points (i.e. $n$).
Existing PCQA methods tend to treat the input point cloud as a whole to extract features, holding the assumption that every region in a point cloud contributes equally to the geometry quality of the entire model \cite{Liuyp22}.
However, those regions (e.g. the region with high curvature) in the point cloud should play a more important role in the final assessment than in other regions \cite{Hual21}.

Firstly, to make the generated patches cover the entire model as much as possible, the FPS (Farthest Point Sampling) strategy \cite{qi2017pointnet++} is employed to sample a fixed number of points from the input degraded point cloud.
These sampled points are used as center points for the next neighboring points selection.
Note that, given paired point clouds, the quality comparison between patches of different point clouds should be performed on the same region.
However, input paired point clouds may have different number of points, since some distortions (e.g. compression sampling) may reduce the number of points.
By deploying center points, the correspondence between patches of paired point clouds can be assured.
In practice, the center points can be sampled from just one of the paired point clouds.

\begin{figure}[!t]
	\centering
	\subfloat[]{\includegraphics[width=0.22\textwidth]{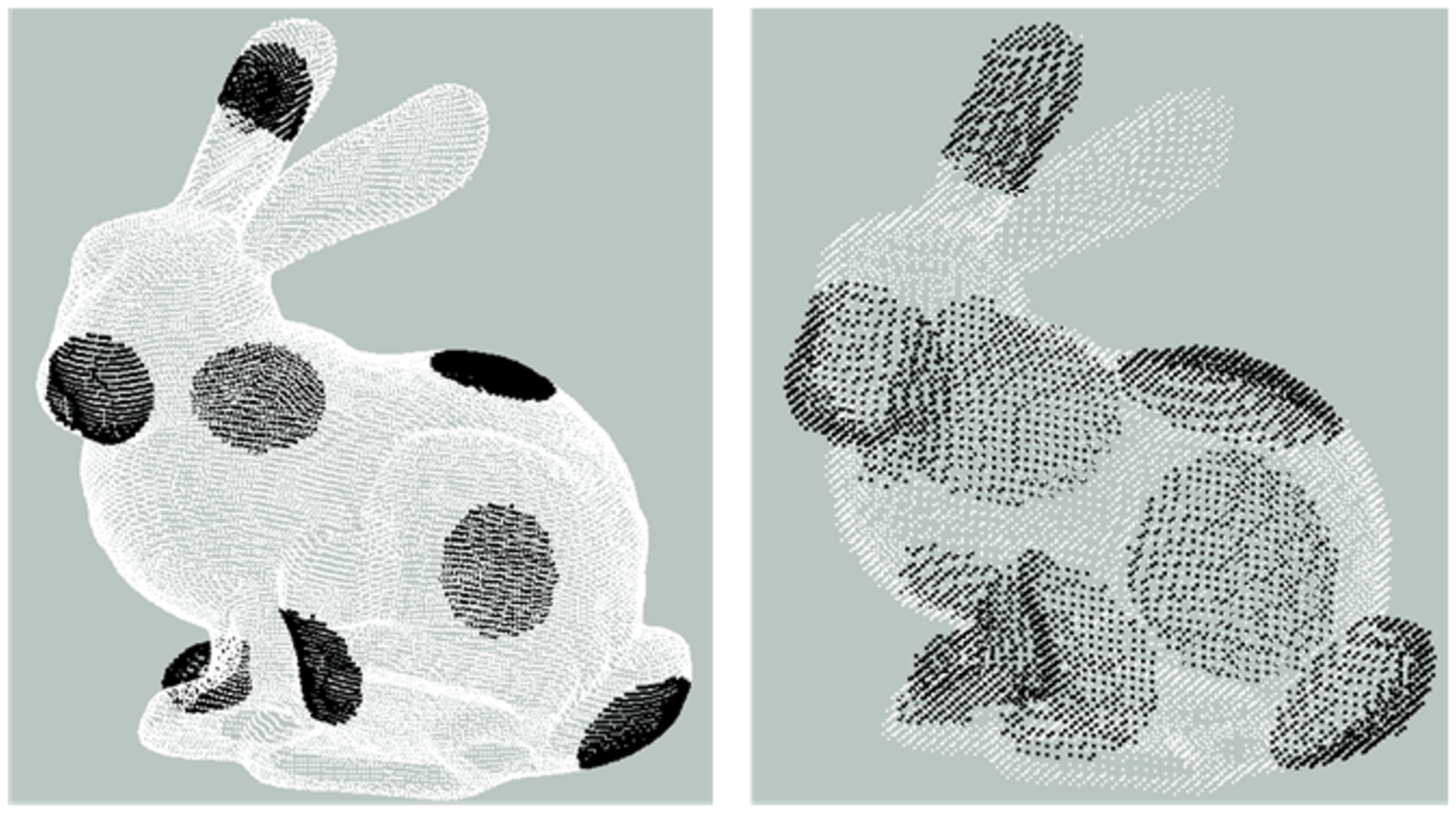}
		\label{fig:sample:a}}
	\subfloat[]{\includegraphics[width=0.22\textwidth]{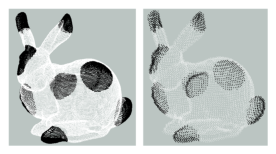}
		\label{fig:sample:b}}
	\caption{Comparison of patch generation between two different neighborhood selection strategies on input paired point clouds with different point numbers. (a) Patch generation based on $k$ nearest neighbors selection. (b) Patch generation based on a fixed radius $r$.}
	\label{fig:sample}
\end{figure}

Then, sampled center points are used to generate patches with a fixed number of points for the input paired point clouds, respectively.
Specifically, for each center point, its neighboring points within a sphere of radius $r$ located at the center point, rather than its $k$ nearest neighbors, are selected to form a patch with $n$ points.
Those patches whose point numbers are less than $n$ are padded to $n$ points by randomly selecting points from their existing points.
Fig. \ref{fig:sample} illustrates the patch generation of input coupled point clouds with different point numbers through the two sampling strategies, respectively.

In addition, for each patch $P_i=\{p_{i1},..., p_{in} | p_{ij} \in \mathbb{R}^3\}$ ($i \in [1, N]$, $j \in [1, n]$) with the center point $c_i$, the absolute coordinates of points in each patch are converted into relative coordinates:
\begin{equation}
	{P_i}^\prime  = \left[ {\begin{array}{*{20}{c}}
			{{p_{i1}}^\prime }\\
			{ \cdots }\\
			{{p_{in}}^\prime }
	\end{array}} \right] = \left[ {\begin{array}{*{20}{c}}
			{p_{i1} - c_i}\\
			{ \cdots }\\
			{p_{in} - c_i}
	\end{array}} \right].
\end{equation}

\subsection{Geometric Feature Extraction}
\begin{figure*}[htbp]
	\centering
	\includegraphics[width=0.8\textwidth]{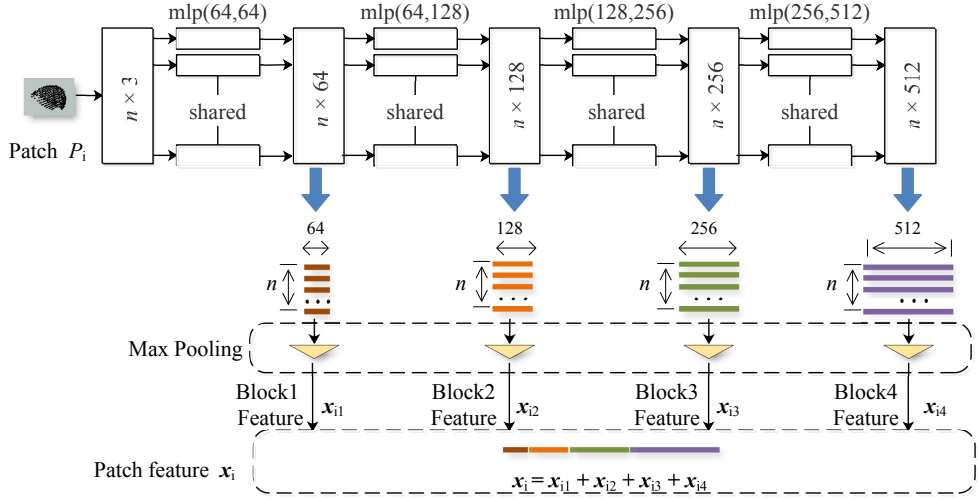}
	\caption{Structure of the geometric feature extraction module.}
	\label{fig:feature_net}
\end{figure*}

The geometric feature extraction module, which consists of four blocks, aims to extract multi-scale point-wise quality-aware geometric features of each patch, as illustrated in Fig. \ref{fig:feature_net}.
Each block is characterized by a shared weight Multi-layer Perceptron (MLP), which is implemented by 1-dimensional convolution.

For each patch $P_i$, the output features of each block are firstly aggregated through the maximum pooling operation,
\begin{equation}
	\left\{ {\begin{array}{*{20}{l}}
			{{\textbf{x}_{i1}} = \mathop {\max }\limits_{j \in [1, n]} ({h_1}({p_{ij}}))}\\
			{{\textbf{x}_{i2}} = \mathop {\max }\limits_{j \in [1, n]} ({h_2}({h_1}({p_{ij}})))}\\
			{{\textbf{x}_{i3}} = \mathop {\max }\limits_{j \in [1, n]} ({h_3}({h_2}({h_1}({p_{ij}}))))}\\
			{{\textbf{x}_{i4}} = \mathop {\max }\limits_{j \in [1, n]} ({h_4}({h_3}({h_2}({h_1}({p_{ij}})))))}
	\end{array}} \right.,
\end{equation}
where ${\textbf{x}_{i1}} \in {\mathbb{R}^{64}}$, ${\textbf{x}_{i2}} \in {\mathbb{R}^{128}}$, ${\textbf{x}_{i3}} \in {\mathbb{R}^{256}}$, ${\textbf{x}_{i4}} \in {\mathbb{R}^{512}}$, $h_1$, $h_2$, $h_3$, and $h_4$ are the four fully connected layers in the shared weight MLP, respectively.
Then, the obtained four feature vectors are concatenated to form the representative hierarchical feature vector $\textbf{x}_{i}$ of the $i$-th patch:
\begin{equation}
	{\textbf{x}_i} = [{\textbf{x}_{i1}}, {\textbf{x}_{i2}}, {\textbf{x}_{i3}}, {\textbf{x}_{i4}}] \in {\mathbb{R}^{960}}.
\end{equation}

\subsection{Weighted Quality Index Calculation}

\begin{figure}[!t]
	\centering
	\includegraphics[width=0.45\textwidth]{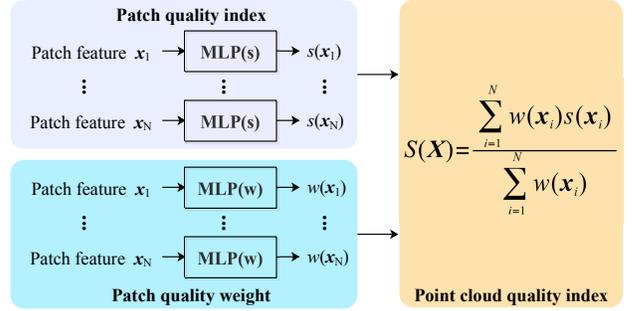}
	\caption{Structure of the weighted quality index calculation module.}
	\label{fig:quality-index}
\end{figure}

After obtaining geometric features $\textbf{x}_{i}$ of each patch $P_i$, a weighted quality calculation module is introduced to calculate the quality index of the whole point cloud, as illustrated in Fig .\ref{fig:quality-index}.

Firstly, a patch quality index prediction network (MLP(s)) is designed to compute the quality index of each patch.
It should be noted that not every patch in a point cloud contributes equally to the geometry quality of the entire model.
Therefore, a patch quality weight prediction network (MLP(w)) is then proposed to assign an adaptive weight to each patch. 
Both MLP(s) and MLP(w) adopt the same MLP model, in which the fully connected layer sizes are 960, 512, 256, and 64, respectively.
The Sigmoid layer is employed to output a single scalar which is indicative of quality index and weight, respectively.
Finally, the overall quality index $S(\textbf{X})$ of the entire point cloud composed of $N$ patches can be obtained by:
\begin{equation}
	S(\textbf{X}) = \frac{{\sum\limits_{i = 1}^N {w({\textbf{x}_i})s({\textbf{x}_i})} }}{{\sum\limits_{i = 1}^N {w({\textbf{x}_i})} }},
\end{equation}
where $s(\textbf{x}_i)$ is the quality index of $P_i$, and $w(\textbf{x}_i)$ is the quality weight of $P_i$.

\subsection{Pairwise Rank Learning}

The pairwise rank learning module aims to learn a function $F$ which takes the overall quality indexes of point cloud $A$ and point cloud $B$ (denoted by $S(\textbf{X}_A)$ and $S(\textbf{X}_B)$, respectively) as inputs and computes the probability that $A$ is better than $B$: $P_{AB} = F(S(\textbf{X}_A), S(\textbf{X}_B))$.
Inspired by the RankNet \cite{Burgesc05,mkd17,Liuxl17,Hub19}, the overall quality indexes $S(\textbf{X}_A)$ and $S(\textbf{X}_B)$ are first subtracted, and then fed through a Sigmoid function to convert it into a probability output:
\begin{equation}
	{P_{AB}} =\frac{{{e^{{S(\textbf{X}_A) - S(\textbf{X}_B)}}}}}{{1 + {e^{{S(\textbf{X}_A) - S(\textbf{X}_B)}}}}},
\end{equation}
where $P_{AB}$ is the modeled posterior $P(S(\textbf{X}_A) > S(\textbf{X}_B))$.
Finally, the cross entropy function is employed as the ranking loss function:
\begin{equation}
	{L_{AB}} =  - {\bar P_{AB}}\log {P_{AB}} - (1 - {\bar P_{AB}})\log (1 - {P_{AB}}),
\end{equation}
where ${\bar P_{AB}}$ is the desired target probability of $P_{AB}$.

\subsection{Fine-tuning for Quality Score Calculation }
\label{sec:Fine-tuning}

The pre-trained GQANet can also be used to predict absolute subjective scores for degraded point clouds after fine-tuning.
Given $M$ point clouds in mini-batch with subjective scores, the GQANet can be fine-tuned using squared Euclidean distance as the loss function in place of the ranking loss used for the Siamese network:
\begin{equation}
	L\left(y_{i}, \hat{y}_{i}\right) = \frac{1}{M} \sum_{i = 1}^{M}\left(y_{i}-\hat{y}_{i}\right)^{2},
\end{equation}
where $y_{i}$ is the ground truth quality score of the $i$-th point cloud, and $\hat{y}_{i}$ is the predicted score from the network. 

\section{Datasets}
\label{sec:Dataset}

\subsection{PRLD Dataset}


To train the proposed PRL-GQA framework, a new large-scale dataset called PRLD is constructed, labeled with different distortion types and degrees.
Currently, only a few small datasets which are publicly accessible have been built in the field of PCQA, such as IRPC \cite{Javaheri21}, CPCD 2.0 \cite{Hual21}, SJTU-PCQA \cite{Yangq21}, SIAT-PCQD \cite{wuxj21}, and WPC \cite{liuq22}.
Among them, the largest dataset WPC contains only 740 degraded samples, including both geometry and color distortions  \cite{liuq22}.
Therefore, it is difficult to employ existing subjective datasets to obtain sufficient ranked pairs for geometry-only quality assessment.
Besides, it is also challenging to collect reliable MOS for a large number of point clouds, since the subjective experiment is time-consuming and expensive. 
Table \ref{Tab:dataset} compares the proposed PRLD dataset with the four existing publicly accessible datasets \cite{liuq22,Liuyp22}.
\begin{table*}[!thp]
	\caption{Comparison of the proposed PRLD dataset and existing publicly accessible PCQA datasets.}
	\label{Tab:dataset}
	\begin{tabular}{lcccc}
		\hline
		Datasets  & Reference samples & Attribute & Distortion type    & Distorted samples \\ \hline
		SIAT-PCQD \cite{wuxj21} & 20         & color     & V-PCC      & 340               \\
		CPCD 2.0 \cite{Hual21}  & 10            & color     & G-PCC, V-PCC, Gassian noise     & 360               \\
		SJTU-PCQA \cite{Yangq21} & 10      & color     & Octree, downsampling, color and geometry noise     & 420               \\
		WPC \cite{liuq22}       & 20                & color     & Gassian noise, dowsampling, G-PCC, V-PCC   & 740               \\ \hline
		\textbf{PRLD} (ours)      & 150            & colorless & \begin{tabular}[c]{@{}c@{}}Gaussian noise, uniform noise, impulse noise, exponential noise, \\ octree-based compression, random dowsampling, grid dowsampling\end{tabular} & 5250              \\ \hline
	\end{tabular}
\end{table*}

\subsubsection{Reference Point Clouds}

\begin{figure*}[htbp]
	\centering
	\includegraphics[width=0.9\textwidth]{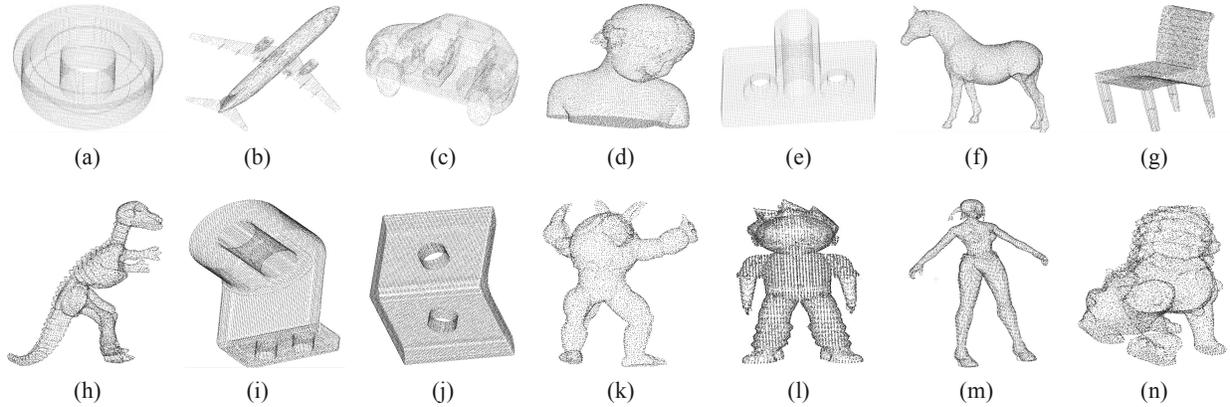}
	\caption{Snapshots of some reference point clouds in the proposed PRLD dataset.}
	\label{fig:samples}
\end{figure*}

The proposed PRLD contains 150 reference point clouds with diverse geometric complexity, including 100 regular-shaped industrial parts and 50 irregular-shaped objects, covering various categories such as vehicles, people, and daily necessities. 
All the reference point clouds come from existing 3D mesh datasets, such as Stanford 3D scanning repository \cite{TurkG94} and ModelNet \cite{Wuzr15}.
For each selected 3D mesh, a uniformly distributed random sampling strategy is employed to obtain the Cartesian coordinates of the crafted point cloud from the surfaces of meshes. 
The number of points of all reference point clouds ranges between 5000 and 50000. 
Fig. \ref{fig:samples} illustrates some reference point clouds. 

\subsubsection{Distortion Generation}

Each reference point cloud is distorted by 7 different types of distortions under 5 distortion levels, including gaussian noise (GN), uniform noise (UN), impulse noise (IN), exponential noise (EN), octree-based compression (OC), random downsampling (RS) and grid downsampling (GS).
To this end, for each reference point cloud, the length of the edge between each point and its nearest neighbor is firstly calculated.
Then, the average length $l_r$  of all edges is employed as the reference value to generate distortions.

\begin{itemize}
	\item Gaussian noise (GN): The function normrnd() in Matlab is used to add the zero-mean Gaussian noise to point positions with standard deviations of {0.1$l_r$, 0.2$l_r$, 0.35$l_r$, 0.5$l_r$, 0.7$l_r$}, respectively.
	
	\item Uniform noise (UN): The function rand() in Matlab is employed to add the zero-mean uniform noise to point positions through randomly offsetting each point along the $x$, $y$, and $z$ direction independently among [-0.3$l_r$, 0.3$l_r$], [-0.6$l_r$, 0.6$l_r$], [-1.05$l_r$, 1.05$l_r$], [-1.5$l_r$, 1.5$l_r$], and [-2.1$l_r$, 2.1$l_r$], respectively.
	
	\item Impulse noise: To add the impulse noise to each point for different distortion levels, the function rand() in Matlab is firstly used to generate random values among [-0.3$l_r$, 0.3$l_r$], [-0.6$l_r$, 0.6$l_r$], [-1.05$l_r$, 1.05$l_r$], [-1.5$l_r$, 1.5$l_r$], and [-2.1$l_r$, 2.1$l_r$], respectively. Then, the threshold value is set to 0.1$l_r$ to produce impulse noise for each point.
	
	\item Exponential noise (EN): The function exprnd() in Matlab is adopted to add exponential noise to each point along the $x$, $y$, and $z$ direction independently with the mean parameter 0.1$l_r$, 0.2$l_r$, 0.35$l_r$, 0.5$l_r$, 0.7$l_r$, respectively.
	
	\item Octree-based compression (OC): The function OctreePointCloudCompression()  provided by the well-known Point Cloud Library (PCL) is used to compress each point cloud by setting the octree resolution at  0.01, 0.0116, 0.014, 0.019 and 0.025 respectively.
	
	\item Random downsampling (RS): The function pcdownsampling()  in Matlab is employed to randomly downsample the point cloud by removing 15\%, 25\%, 40\%, 55\%, and 70\% points from the original point, respectively.
	
	\item Grid downsampling (GS): The function pcdownsampling() in Matlab is used to downsample the point cloud through a sampling grid with the resolution 1.2$l_r$, 1.4$l_r$, 1.65$l_r$, 2.0$l_r$, and 2.5$l_r$, respectively.
\end{itemize}

\subsubsection{Paired Samples}

For each reference point cloud, each type of distortion can lead to 5 different levels of degraded versions.
Thus, there are $C(6, 2) = \frac{6!}{2!(6-2)!} = 15$ pairs of training samples for each type of distortion, and $15 \times 7 = 105$ pairs of training samples for each reference point cloud.
Each pair of training samples is a pairwise comparison consisting of two versions of the same reference point cloud (e.g., $A'$ and $A''$), along with a label ${\bar P_{A'A''}}$, which is the target probability that point cloud $A'$ is better than point cloud $A''$.
Meanwhile, each pair of training samples only suffer from a single type of distortion. 
Note that all the generated paired samples have no ground-truth scores yet contain quality ranking information.

\subsection{PCGD-PMOS Dataset}


To fine-tune the pre-trained GQANet to predict absolute quality scores, a large-scale quality-annotated dataset containing 5250 geometrically distorted samples with pseudo-MOS, called PCGD-PMOS, is established.
Existing publicly available PCQA datasets, such as IRPC \cite{Javaheri21}, CPCD 2.0 \cite{Hual21}, SJTU-PCQA \cite{Yangq21}, SIAT-PCQD \cite{wuxj21}, and WPC \cite{liuq22}, are usually too small and contain only a small number of colorless point clouds.
For example, the largest dataset, i.e., WPC \cite{liuq22}, only possesses 60 geometrically distorted samples which are derived from 20 pristine color and colorless point clouds.
Furthermore, collecting MOS for point clouds by subjective experiment is laborious and usually needs highly controlled conditions. 
Therefore, it is a feasible way to build the dataset PCGD-PMOS using pseudo-MOS, which has been successfully applied in point clouds \cite{Alexioue18icme, Liuyp22} and images \cite{Wujj20,oufz21}.

In this paper, the angular similarity metric proposed by Alexiou et al. \cite{Alexioue18icme} is employed to calculate pseudo-MOS.
As a kind of plane-to-plane distance, the performance of the referred metric has been demonstrated to be consistent with subjective quality assessment scores under certain types of geometry distortions, such as Gaussian noise, and compression-like artifacts \cite{Alexioue18icme}.
Specifically, for each reference point cloud and its degraded versions in the PRLD Dataset, the degree of distortion is measured by the angular similarity of the reference point cloud and its distorted version and then is directly used as a subjective score of [0.0, 1.0].
Finally, the PCGD-PMOS dataset containing $150 \times 7 \times 5 = 5250$ geometrically distorted point cloud samples with pseudo-MOS can be obtained for fine-tuning.

\section{Experimental Results and Analysis}
\label{sec:Experimental}

\subsection{Evaluation Metrics}


Three common evaluation metrics including Pearson linear correlation coefficient (PLCC) \cite{Sedgwicke12}, Kendall’s rank-order correlation coefficient (KRCC) \cite{liuq21tcsvt} , and Spearman's rank-order correlation coefficient (SRCC) \cite{Yangq21} are employed to evaluate the performance of PRL-GQA.
Given $N$ distorted point clouds, the ground truth of the $i$-th point cloud is denoted by $y_i$, and the predicted quality score is $\hat{y}_{i}$.
The PLCC measures the linear correlation between the ground truth and predicted quality scores.
Formally, PLCC is defined as:
\begin{equation}
	\label{eq:PLCC}
	\mathrm{PLCC} = \frac{\sum_{i = 1}^{N}\left(y_{i}-\bar{y}\right)\left(\hat{y}_{i}-\overline{\hat{y}}\right)}{\sqrt{\sum_{i}^{N}\left(y_{i}-\bar{y}\right)^{2}} \sqrt{\sum_{i}^{N}\left(\hat{y}_{i}-\overline{\hat{y}}\right)^{2}}},
\end{equation}
where $\bar{y}$ and $\overline{\hat{y}}$ are the arithmetic means of the ground truth and predicted quality scores, respectively.
The KRCC evaluates the association between two ordinal (two ranked variables, not necessarily intervals) variables via
\begin{equation}
	\label{eq:KRCC}
	\mathrm{KRCC} = \frac{{2({N_c} - {N_d})}}{{N (N - 1)}},
\end{equation}
where $N_c$ and $N_d$ are the number of concordant (of consistent rank order)  and discordant (of inconsistent rank order) pairs, respectively.
The SRCC assesses the monotony between the ground truth and predicted quality scores, which is computed as
\begin{equation}
	\label{eq:SRCC}
	\mathrm{SRCC} = 1 - \frac{6}{{N({N^2} - 1)}}\sum\limits_{i = 1}^N {{{\left( {v_i - u_i} \right)}^2}},
\end{equation}
where $v_i$ is the rank of $y_i$ in the ground truth scores, $u_i$ is the rank of $\overline{\hat{y}}$ in predicted quality scores.

The ranking accuracy of the proposed PRL-GQA is evaluated through Eq. (\ref{eq:Acc}) ,
\begin{equation}
	\label{eq:Acc}
	\mathrm{Acc} = \frac{TP + TN}{TP + TN + FP + FN},
\end{equation}
where $TP$ (True Positive), $TN$ (True Negative), $FP$ (False Positive), and $FN$ (False Negative) denote the number of predictions where the PRL-GQA correctly predicts positive cases as positive, the number of predictions where the PRL-GQA correctly predicts negative cases as negative, the number of predictions where the PRL-GQA incorrectly predicts negative cases as positive, and the number of predictions where the PRL-GQA incorrectly predicts positive cases as negative, respectively. 

In addition, the list-wise ranking consistency test $L_{Test}$ \cite{Makd17} is introduced to evaluate the ranking consistency between the distortion levels and the model predictions.
The goal of $L_{Test}$ is to evaluate the robustness of PRL-GQA when rating point clouds with the same content and the same distortion type but different distortion levels, with the assumption that the quality of a point cloud degrades monotonically with the increase of the distortion level for any distortion type. 
Given a dataset with $M$ pristine point clouds, $K$ distortion types, and $L$ distortion levels, $L_{Test}$ is defined as
\begin{equation}
	\label{eq:rankcon}
	L_{Test} = \frac{1}{MK}\sum_{i=1}^{M}\sum_{j=1}^{K}\mathrm{SRCC} \left (  l_{ij}, q_{ij}\right ), 
\end{equation}
where $ l_{ij}$ and $q_{ij}$ are both length-$L$ vectors representing the distortion levels and the corresponding quality scores, respectively.

\subsection{Experimental settings}

\textbf{Training and testing datasets}. 
To evaluate the performance of PRL-GQA, the reference point clouds in the PRLD dataset are randomly divided in a way that 80\% of the samples are used for training while the remaining 20\% are for testing.
Overall, there are a total of 12600 pairwise samples for training and 3150 paired samples for testing. 
The proposed PCGD-PMOS dataset with pseudo-MOS is employed to assess the performance of GQANet, where the reference point clouds are randomly split into two groups: 80\% for fine-tuning the pre-trained GQANet and 20\% for testing.

\textbf{Parameters setting}.
In the training stage, the farthest point sampling (FPS) \cite{qi2017pointnet++} and sphere matching are used to sample 64 overlapping patches with 512 points for each point cloud. 
The Adam optimization with a batch of 4 is adopted for training.
The initial learning rate is $10^{-5}$ and decreased by a factor of 0.5 every 2 epochs for a total of 20 epochs.

\begin{figure}[!htbp]
	\centering
	\includegraphics[width=0.5\textwidth]{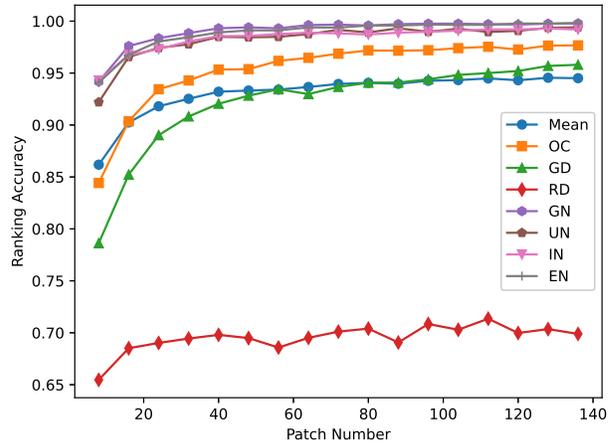}
	\caption{Results of the ranking accuracy of PRL-GQA under different patch numbers.}
	\label{fig:number_patch}
\end{figure}

In the testing stage, the patch number $N$ used in the patch generation module is set to 112 in this paper.
Specifically, it is determined experimentally by comparing the ranking accuracy under different patch numbers on the PRLD dataset.
Fig. \ref{fig:number_patch} illustrates the experimental results, which indicates that, with the increase of $N$, the ranking accuracy under different distortions increases until it is stable. 
Thus, to balance the efficiency and accuracy, the patch number $N$ is set to 112 in the following experiments.

\begin{table*}[!ht]
	\centering
	\caption{Ranking accuracy of testing paired samples (\%)}
	\label{tab:pair_results}
		\begin{tabular}{clccccccc|c}
			\hline	 
			Types                  & \multicolumn{1}{c}{Method} & OC        & GS          & RS        & GN        & UN        & IN        & EN        & Mean         \\ \hline
			\multirow{9}{*}{Full-Reference} & Po2Po$_{MSE}$  \cite{Rufaelm16}            & 99.330        & 100.000       & 100.000       & 100.000       & 100.000       & 100.000       & 100.000       & \textbf{99.900 }       \\
			& Po2Po$_{Hausdorff}$ \cite{Rufaelm16}       & \textbf{100.000}       & \textbf{100.000}       & 98.890        & 100.000       & 100.000       & \textbf{100.000}       & 99.110        & 99.710        \\
			& Po2Po$_{PSNR}$  \cite{Rufaelm16}           & 99.330        & 100.000       & 100.000       & 100.000       & 100.000       & 100.000       & 100.000       & 99.900        \\
			& Po2PL$_{MSE}$ \cite{Tiand17}             & 98.670        & 98.890        & 100.000       & 100.000       & 100.000       & 100.000       & 100.000       & 99.650        \\
			& Po2PL$_{Hausdorff}$ \cite{Tiand17}        & 100.000       & 94.890        & 97.560        & 100.000       & 100.000       & 100.000       & 99.110        & 98.790        \\
			& Po2PL$_{PSNR}$  \cite{Tiand17}           & 98.670        & 98.890        & 100.000       & 100.000       & 100.000       & 100.000       & 100.000       & 99.650        \\
			& PL2PL$_{MSE}$ \cite{Alexioue18icme}             & 92.670        & 99.780        & \textbf{100.000}       & \textbf{100.000}       & \textbf{100.000}       & 89.330        & \textbf{100.000}       & 97.400        \\
			& PL2PL$_{Hausdorff}$ \cite{Alexioue18icme}             & 92.670        & 79.780        & 84.220        & 92.220        & 89.330        & 87.560        & 91.330        & 88.160        \\
			& PL2PL$_{PSNR}$ \cite{Alexioue18icme}            & 95.110        & 99.780        & 100.000       & 100.000       & 100.000       & 91.330        & 100.000       & 98.030        \\ \hline
			No-Reference       & PRL-GQA (ours)   &97.533 &95.000 &71.356 &99.711 &98.956 &99.200 &99.644 &94.486 \\  \hline
		\end{tabular}	
\end{table*}

\textbf{Compared methods}.
As discussed in Section \ref{sec:Related Works}, to the best of our knowledge, the PRL-GQA is the first point-based no-reference geometry-only quality assessment method. 
Existing no-reference PCQA metrics are all designed to assess point clouds with both geometry and color attributes \cite{Yangq22,Liuyp22}.
While the proposed PRL-GQA framework aims to conduct objective geometry-only quality assessment.
Therefore, no direct comparisons can be made with existing no-reference PCQA metrics.
For fair comparison and analysis, three full-reference objective assessment metrics including Point-to-Point (Po2Po) \cite{Rufaelm16}, Point-to-Plane (Po2PL) \cite{Tiand17} and Plane-to-Plane (PL2PL) \cite{Alexioue18icme} are employed to evaluate geometry-only quality in this paper.
And, three pooling strategies including Mean Squared Error (MSE), Hausdorff distance, and geometric PSNR are introduced to obtain overall quality distortion \cite{Javaheri21, zhoul22}.
In total, 9 point-based metrics are tested on the proposed datasets, e.g., Po2Po$_{MSE}$, Po2Po$_{Hausdorff}$, Po2Po$_{PSNR}$, Po2PL$_{MSE}$ , Po2PL$_{Hausdorff}$, Po2PL$_{PSNR}$, PL2PL$_{MSE}$, PL2PL$_{Hausdorff}$, PL2PL$_{PSNR}$.

\subsection{Quality Ranking}

To evaluate the ranking accuracy of the proposed PRL-GQA, all the compared methods first calculate the quality indexes of point clouds in each testing pair, and then determine their ranking order according to predicted indexes. 
Table \ref{tab:pair_results} shows the ranking accuracy for each type of distortion on the entire testing dataset.
From Table \ref{tab:pair_results}, it can be seen that all the full-reference methods perform well for all distortion types, especially the four types of noise distortion.
As a no-reference geometry quality assessment method, the proposed PRL-GQA performs a little worse than the three full-reference methods.
To be specific, in terms of RS distortion, the ranking accuracy of the proposed PRL-GQA is only 71.356\%.
The reason may be that the random downsampling operation may result in uneven distribution of distortion in the whole point cloud. 
Consequently, the distortion of each patch obtained during each test may be inconsistent, leading to the incorrect ranking order.
However, for the other distortion types, the proposed PRL-GQA achieves comparative performance among all compared methods, and the ranking accuracy is more than 95\%.

To further verify the generalization of PRL-GQA, each reference point cloud for testing in the PRLD dataset is degraded by 7 distortion types with 30 distortion levels.
The output of the GQANet without fine-tuning is directly used as the predicted distortion level.
The list-wise ranking consistency is then employed to evaluate the performance of the proposed PRL-GQA.
Table \ref{tab:rank_consistency} lists the SRCC coefficient as well as the ranking consistency between predicted levels and ground-truth levels under various distortions.
It can be observed that the proposed GQANet, which is trained on the testing datasets with only 5 distortion levels, can still obtain high list consistency for each type of distortion, except for the RS distortion.
Therefore, the proposed PRL-GQA is qualified for assessing or ranking degraded point clouds with more subtle differences.

\begin{table*}[!ht]
	\centering
	\caption{SRCC coefficients and list-wise ranking consistency results under 30 distortion levels.}	
	\label{tab:rank_consistency}
		\begin{tabular}{clccccccc|c}
			\hline	 
			Types                  & \multicolumn{1}{c}{Method} & OC        & GS          & RS        & GN        & UN        & IN        & EN        & $L_{Test}$        \\ \hline
			\multirow{9}{*}{Full-Reference} & Po2Po$_{MSE}$ \cite{Rufaelm16}     & \textbf{1.000}        & \textbf{1.000}       & \textbf{1.000}          & \textbf{1.000}       & \textbf{1.000}       & \textbf{1.000 }      & 0.999        & \textbf{1.000}        \\
			& Po2Po$_{Hausdorff}$ \cite{Rufaelm16}       & 1.000       & 0.990       & 0.953        & 0.981       & 0.999       & 1.000       & 0.981        & 0.986        \\
			& Po2Po$_{PSNR}$  \cite{Rufaelm16}           & 1.000        & 1.000       & 1.000          & 1.000       & 1.000       & 1.000       & 0.999        & 1.000        \\
			& Po2PL$_{MSE}$ \cite{Tiand17}             & 1.000        & 1.000        & 0.999       & 1.000       & 1.000       & 1.000       & \textbf{1.000}       & 1.000        \\
			& Po2PL$_{Hausdorff}$ \cite{Tiand17}       & 0.998       & 0.973        & 0.904        & 0.984       & 0.999       & 1.000       & 0.982        & 0.977       \\
			& Po2PL$_{PSNR}$  \cite{Tiand17}            & 1.000        & 1.000        & 0.999       & 1.000       & 1.000       & 1.000       & 1.000       & 1.000        \\
			& PL2PL$_{MSE}$  \cite{Alexioue18icme}     & 0.993        & 1.000        & 1.000       & 1.000       & 0.999       & 0.997        & 1.000       & 0.998        \\
			& PL2PL$_{Hausdorff}$ \cite{Alexioue18icme}   & 0.419        & 0.400        & 0.627        & 0.890        & 0.725        & 0.723        & 0.863        & 0.664        \\
			& PL2PL$_{PSNR}$ \cite{Alexioue18icme}       & 0.989        & 1.000        & 1.000       & 1.000       & 0.999       & 0.998        & 0.999       & 0.998        \\ \hline
			No-Reference   &PRL-GQA (ours)                    &0.701 &0.959 &0.097 &0.737 &0.899 &0.874 &0.784 & 0.722  \\ \hline  
	\end{tabular}
\end{table*}

\subsection{Quality Score Calculation}

The pre-trained GQANet also possess the capacity to calculate absolute quality scores.
To evaluate its predicting performance, firstly, the pre-trained GQANet is directly tested on the G-PCD \cite{Alexioue17} dataset without fine-tuning.
The G-PCD \cite{Alexioue17} is a small dataset with subjective ground-truth MOS, containing 5 colorless reference point clouds and 40 samples distorted by octree-puring and Gaussian noise \cite{Alexioue17}.
After that, the pre-trained GQANet is fine-tuned on the training set and tested on the testing set of the PCGD-PMOS dataset.
The output of GQANet is directly used as absolute quality scores.
PLCC, SRCC, and KRCC are used to compare the performance of the GQANet with existing full-reference geometry metrics. 
Comparison results are shown in Table \ref{tab:G-PCDresult} and Table \ref{tab:PCGDresult}, respectively.
From Table \ref{tab:G-PCDresult}, the results show that the pre-trained GQANet without fine-tuning can still obtain comparative predicting performance.
Table \ref{tab:PCGDresult} further demonstrates that, after fine-tuning on a dataset with MOS data, the pre-trained no-reference GQANet achieves competitive results and even outperforms some full-reference evaluation metrics.

\begin{table}[!t]
	\centering
	\caption{Absolute quality score prediction performance of the GQANet on the G-PCD \cite{Alexioue17} dataset without fine-tuning.}	
	\label{tab:G-PCDresult}
		\begin{tabular}{clccc}
			\hline	
			Types                  & \multicolumn{1}{c}{Method} & PLCC        & SRCC          & KRCC         \\ \hline
			\multirow{9}{*}{Full-Reference} & Po2Po$_{MSE}$ \cite{Rufaelm16}             & 0.775        & 0.738       &  0.646        \\
			& Po2Po$_{Hausdorff}$ \cite{Rufaelm16}       & 0.818       &0.762       & 0.597       \\
			& Po2Po$_{PSNR}$  \cite{Rufaelm16}           & 0.786        & 0.774       & 0.682        \\
			& Po2PL$_{MSE}$ \cite{Tiand17}             & 0.795        & 0.761        & 0.661        \\
			& Po2PL$_{Hausdorff}$ \cite{Tiand17}       & 0.818       & 0.756       & 0.589       \\
			& Po2PL$_{PSNR}$  \cite{Tiand17}            & 0.797        & 0.779        & 0.669        \\
			& PL2PL$_{MSE}$ \cite{Alexioue18icme}             &\textbf{0.904}       &\textbf{0.906}       &\textbf{0.744}   \\
			& PL2PL$_{Hausdorff}$ \cite{Alexioue18icme}             & 0.419        & 0.400        & 0.627  \\
			& PL2PL$_{PSNR}$  \cite{Alexioue18icme}           & 0.690        & 0.654        & 0.615      \\ \hline
			No-Reference     & GQANet (ours)                    & 0.624  & 0.613  & 0.430 \\ \hline  
	\end{tabular}
\end{table}

\begin{table}[!t]
	\centering
	\caption{Absolute quality score prediction performance of the GQANet on the PCGD-PMOS dataset with pseudo-MOS.}	
	\label{tab:PCGDresult}
		\begin{tabular}{clccc}
			\hline	 
			Types                  & \multicolumn{1}{c}{Method} & PLCC        & SRCC          & KRCC         \\ \hline
			\multirow{9}{*}{Full-Reference} & Po2Po$_{MSE}$ \cite{Rufaelm16}             &0.507       & 0.632       &0.461        \\
			& Po2Po$_{Hausdorff}$ \cite{Rufaelm16}       &0.395       &0.731       &0.528       \\
			& Po2Po$_{PSNR}$ \cite{Rufaelm16}            &0.422        &0.748       &0.530        \\
			& Po2PL$_{MSE}$ \cite{Tiand17}             &0.711        &0.903        &0.727        \\
			& Po2PL$_{Hausdorff}$ \cite{Tiand17}       &0.421       &0.760       &0.558       \\
			& Po2PL$_{PSNR}$ \cite{Tiand17}             &0.498        &0.804        &0.615        \\
			& PL2PL$_{MSE}$  \cite{Alexioue18icme}            &\textbf{0.994}       &\textbf{0.992}    &\textbf{0.931}     \\
			& PL2PL$_{Hausdorff}$  \cite{Alexioue18icme}            &0.607        &0.568        &0.439  \\
			& PL2PL$_{PSNR}$ \cite{Alexioue18icme}            &0.457        &0.374        &0.258      \\ \hline
			No-Reference                  & GQANet (ours)     &0.752  &0.746 & 0.545 \\ \hline  
	\end{tabular}
\end{table}

\subsection{Ablation Experiments}

Three comparative experiments are designed to verify the contribution of each component in the proposed network in this section.


%

\subsubsection{Patch Generation} 

To demonstrate the effectiveness of the patch generation module, the PRL-GQA employs only one geometric feature extraction module which directly takes the entire input point cloud as input to extract feature vectors.
The ranking accuracy of the PRL-GQA without the patch generation module is shown in Table \ref{tab:Ablation_results}.
It can be seen from Table \ref{tab:Ablation_results} that the PRL-GQA with the patch generation module greatly improves the performance compared to the PRL-GQA without the patch generation module. 

\begin{table*}[!htb]
	\centering
	\caption{Ranking accuracy of PRL-GQA with different network structures (\%).  }
	\label{tab:Ablation_results}
		\begin{tabular}{lccccccc|c}
			\hline	 
			Method          & OC         & GS           & RS            & GN            & UN            & IN          & EN           & Mean         \\ \hline	
			PRL-GQA (without patch)          & 84.00          & 24.44         &31.55          & 95.11         & 96.44          & 96.88        & 94.66 & 74.73 \\ 
			PRL-GQA  (without weight)          & 65.13          & 80.02         &67.04          & 80.73         & 83.73          & 83.60         & 83.71 & 77.71 \\ 
			PRL-GQA  (with DGCNN as backbone)          & 88.33          & 88.00         &\textbf{72.76}          & 83.47         & 64.22          & 80.51         & 90.20 & 81.07 \\  \hline
			PRL-GQA (ours)       & \textbf{97.53} &\textbf{95.00} & 71.36 & \textbf{99.71} &\textbf{98.96} & \textbf{99.20} & \textbf{99.64} & \textbf{94.49} \\\hline 
		\end{tabular}
\end{table*}

\subsubsection{Geometric Feature Extraction} 

To capture multi-scale and quality-aware features, the proposed geometric feature extraction module concatenates the hierarchical features extracted from four blocks (Block1 to 4), as illustrated in Fig. \ref{fig:feature_net}
In order to evaluate the effectiveness of four blocks, each block and their combinations are tested.
The ranking accuracy results are shown in Table \ref{tab:module_results}. 
Obviously, with the increase of network depth, the ranking accuracy under various distortion cases rises steadily.
And, these results demonstrate that the combination of four blocks achieves the best overall performance.

\begin{table*}[!ht]
	\centering
	\caption{Ranking accuracy using the geometric feature extraction module with different block and their combinations (\%). }
	\label{tab:module_results}
		\begin{tabular}{lccccccc|c}
			\hline	 
			Component          & OC         & GS           & RS            & GN            & UN            & IN          & EN           & Mean         \\ \hline	
			Block 1        & 72.400          & 77.222          & 61.711         & 95.244          & 96.489          & 96.089         & 96.289 & 85.063\\
			Block 2        & 42.289         & 76.178         & 61.244         & 96.289         & 96.400          & 96.822         & 95.267 & 80.641           \\
			Block 3        & 81.289           & 92.422          & 69.156         & 98.644       & 99.355          & 99.111       & 99.178 & 91.308\\
			Block 4        & 93.800      & 90.844         & 69.511      & 99.533     & 98.933       & 98.022      & 97.511 & 92.594\\
			Block 3+4       & 82.511        & \textbf{99.267}          & 69.578       & 96.533       & 96.999        & 98.400        & 99.156  & 91.778\\
			Block 2+3+4      & 85.911          & 94.978      & 68.133      & \textbf{99.756}          & \textbf{99.911}          & \textbf{99.911}          & \textbf{99.822} & 92.632\\ \hline
			Block 1+2+3+4 (ours)     & \textbf{97.533}          & 95.000       & \textbf{71.356}          & 99.711       & 98.956     & 99.200      & 99.644  & \textbf{94.486} \\ 
			\hline  
		\end{tabular}
\end{table*}

To further elaborate the effectiveness of the geometric feature extraction module, the dynamic graph convolution network (DGCNN) \cite{Wangy19} is used as the feature extraction backbone.
The DGCNN is composed of four Edge Convolution layers and one Multilayer Perception(MLP) layer. 
The output features of four Edge Convolution layers are concatenated to generate the 512 $\times$ 1 features, which are then fed into MLP to obtain the 960 $\times$ 1 features by max pooling.
The ranking accuracy of the modified DGCNN as the feature extraction backbone is shown in Table \ref{tab:Ablation_results}. 
It can be observed that the DGCNN can also effectively extract the features related to the quality of point clouds, and the final ranking accuracy is more than 80\%.
However, it performs poorer than the geometric feature extraction module proposed in this paper.

\subsubsection{Weighted Quality Index Calculation} 

To verify the effectiveness of the weighted quality index calculation module, the PRL-GQA assigns equal weights to all patches which indicate that every patch in the input point cloud contributes equally to the geometry quality of the whole model.
The ranking accuracy of the PRL-GQA without the weighted quality index calculation module is listed in Table \ref{tab:Ablation_results}.
It can be observed that the weighted quality index calculation module improves significantly the ranking accuracy.

\section{Conclusion}
\label{sec:Conclusion}

This paper presents a no-reference PRL-GQA framework for point cloud geometry-only quality assessment.
To mitigate the problem of small point cloud geometry quality assessment datasets and the shortcoming of subjective quality scores, the pairwise rank learning is introduced to learn to rate degraded point clouds.
A large-scale geometry quality assessment dataset, called PRLD, is constructed to generate ranked samples to train the proposed PRL-GQA.
The pre-trained geometry quality assessment network, named GQANet, can also be employed to predict absolute quality scores after fine-tuning on a newly built large-scale quality-annotated dataset PCGD-PMOS with pseudo-MOS.
Experimental results have demonstrated the efficiency of the proposed no-reference PRL-GQA which achieves competitive performance compared with existing full reference geometry quality assessment metrics.



%
\bibliographystyle{IEEEtran}
\bibliography{PA}


\begin{IEEEbiography}[{\includegraphics[width=1in,height=1.25in,clip,keepaspectratio]{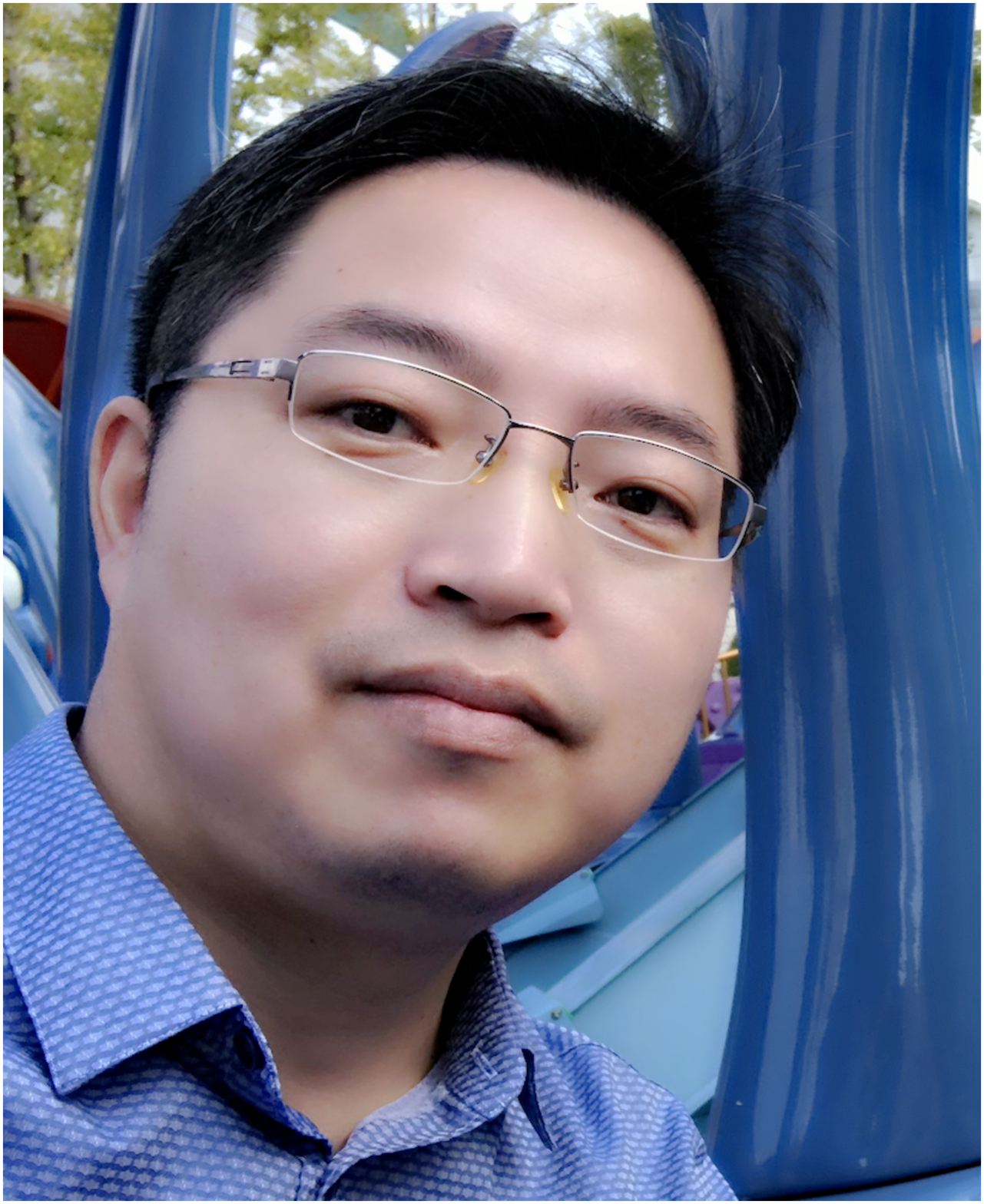}}]{Zhiyong Su}
	is currently an associate professor at the School of Automation, Nanjing University of Science and Technology, China. He received the B.S. and M.S. degrees from the School of Computer Science and Technology, Nanjing University of Science and Technology in 2004 and 2006, respectively, and received the Ph.D. from the Institute of Computing Technology, Chinese Academy of Sciences in 2009. His current interests include computer graphics, computer vision, augmented reality, and machine learning.  
\end{IEEEbiography}

\begin{IEEEbiography}[{\includegraphics[width=1in,height=1.25in,clip,keepaspectratio]{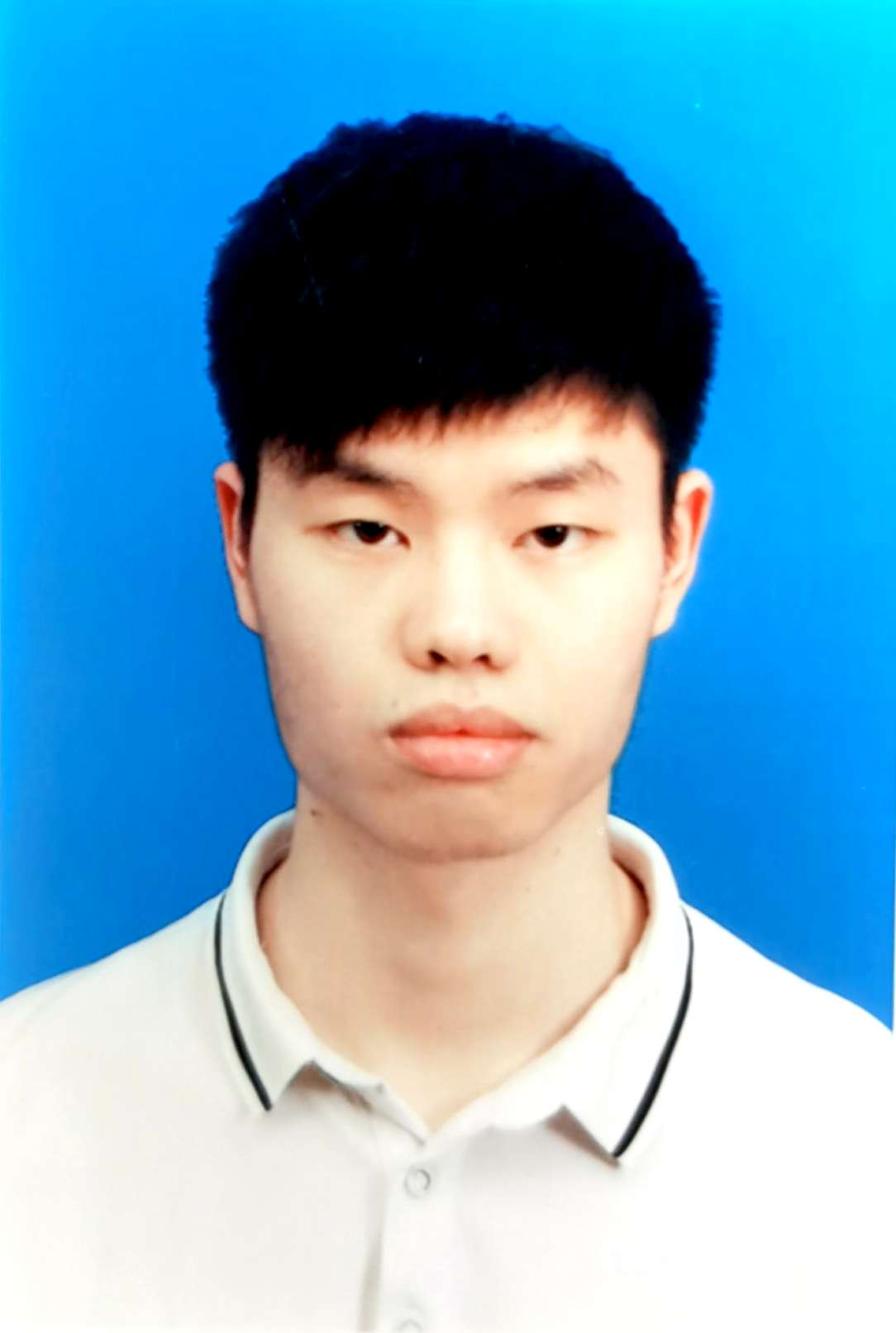}}]{Chao Chu}
	is currently working toward the M.S. degree at the School of Automation, Nanjing University of Science and Technology, China. He received the B.S. degree from the School of mechanical and electronic engineering, Nanjing Forestry University, in 2019. His research interests include computer graphics, quality assessment and machine learning.
\end{IEEEbiography}

\begin{IEEEbiography}[{\includegraphics[width=1in,height=1.25in,clip,keepaspectratio]{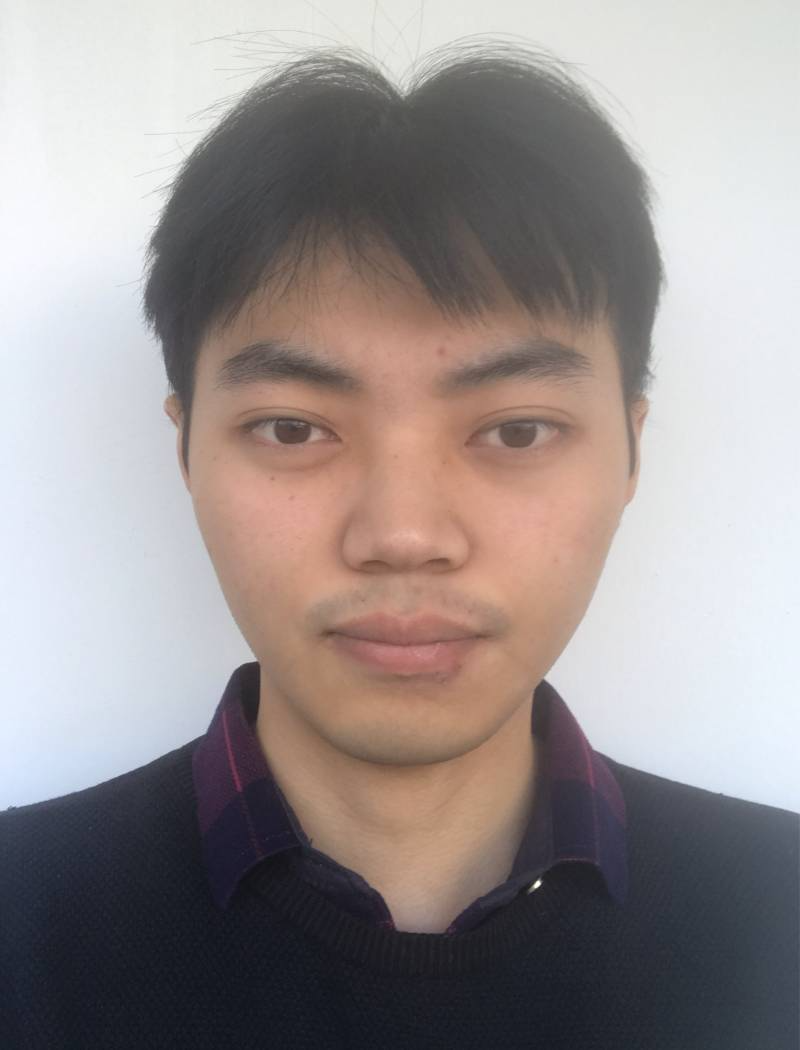}}]{Long Chen}
	received the B.S. degree from Nanjing Normal University, Jiangsu, China in 2021 and now studies at Nanjing University of Science and Technology. His research interests include quality assessment of point cloud.
\end{IEEEbiography}

\begin{IEEEbiography}[{\includegraphics[width=1in,height=1.25in,clip,keepaspectratio]{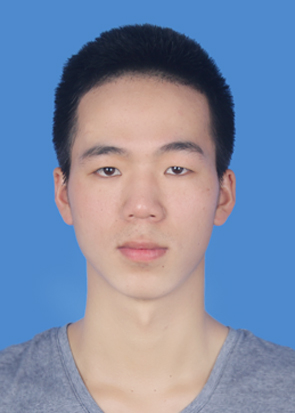}}]{Yong Li}
	received the M.S. and B.S. degrees from the School of Automation, Nanjing University of Science and Technology, China in 2019 and 2022, respectively. His research interests include quality assessment of point clouds and machine learning.
\end{IEEEbiography}

\begin{IEEEbiography}[{\includegraphics[width=1in,height=1.25in,clip,keepaspectratio]{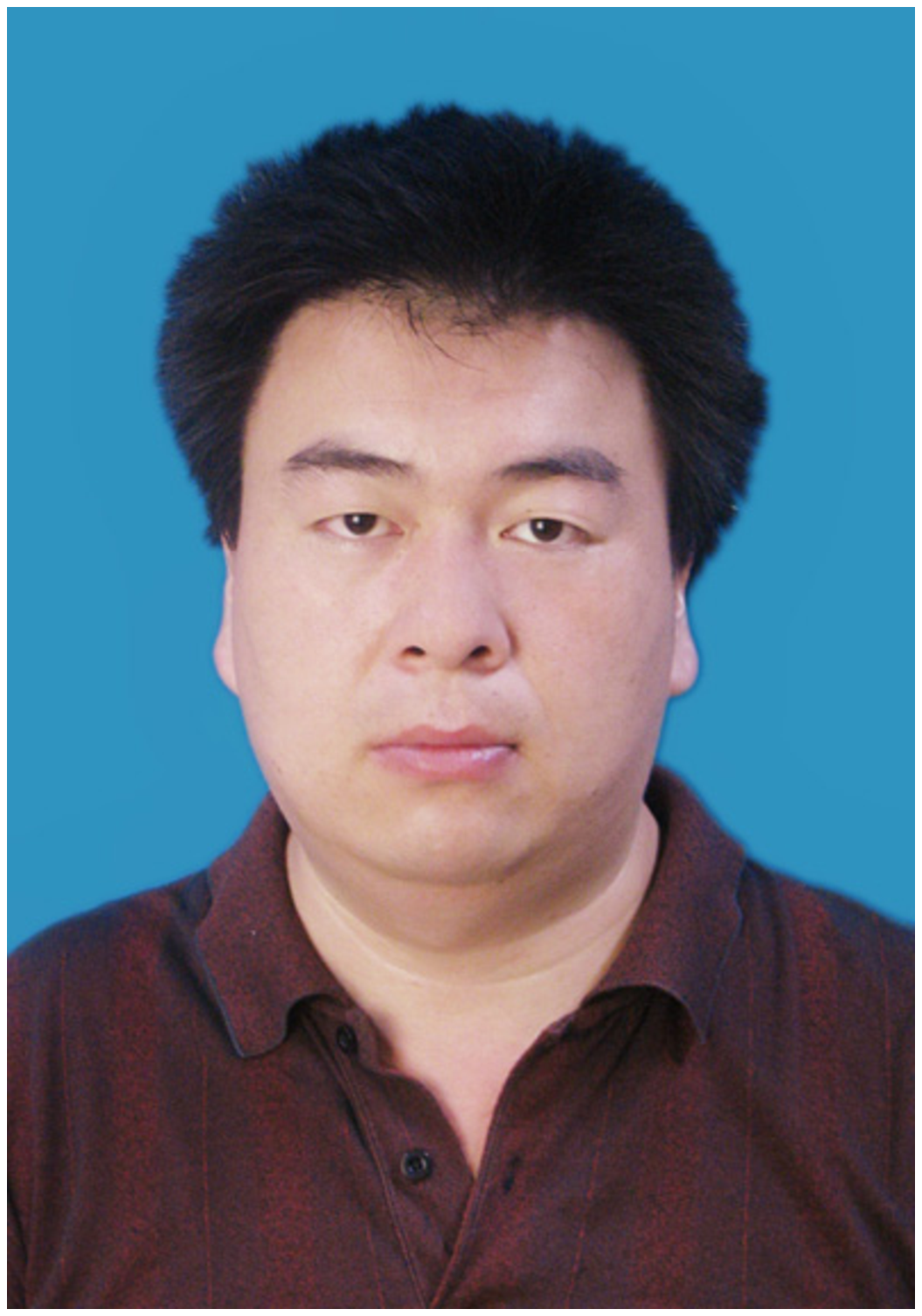}}]{Weiqing Li}
	is currently an associate professor at the School of Computer Science and Engineering, Nanjing University of Science and Technology, China. He received the B.S. and Ph.D. degrees from the School of Computer Sciences and Engineering, Nanjing University of Science and Technology in 1997 and 2007, respectively. His current interests include computer graphics and virtual reality.
\end{IEEEbiography}

\vfill

\end{document}